\documentclass[letter]{aa}

\usepackage{float}
\usepackage{graphicx}
\usepackage{txfonts}
\usepackage{lipsum}
\usepackage{xcolor}
\usepackage{subcaption}         
\usepackage{lscape}             
\usepackage{placeins}           

\newcommand{\fnl}[0]{$f_{\rm NL}$\;}
\newcommand{\fnleqn}[0]{f_{\rm NL}}

\newcommand{\lcdm}{$\Lambda$CDM\xspace}

\usepackage[colorlinks,linkcolor=blue,citecolor=blue,urlcolor=blue ]{hyperref}
                                
\begin{document}

   \title{Unclustered tracers remain unclustered}
   \subtitle{The lack of primordial non-Gaussianity response of bias-zero tracers}


   \author{Celia Merino \inst{1}
        \and Santiago Avila \inst{1}\fnmsep\thanks{santiagoj.avila@ciemat.es}
        \and A. G. Adame \inst{2,3}
        \and A. Anguren \inst{2}
        \and V. Gonzalez-Perez \inst{2,4}
        \and J. Meneses-Rizo \inst{5}
        }

    \institute{Centro de Investigaciones Energ\'eticas, Medioambientales y Tecnol\'ogicas (CIEMAT), Madrid, Spain 
    \and
    Departamento de F\'isica Te\'orica,  Universidad Aut\'onoma de Madrid, 28049 Madrid, Spain
    \and 
    Department of Astrophysics, University of Vienna, Türkenschanzstrasse 17, 1180 Vienna, Austria 
    \and 
    Centro de Investigación Avanzada en Física Fundamental (CIAFF), Universidad Autónoma de Madrid, ES-28049 Madrid, Spain
    \and
    Instituto de Física, Universidad Nacional Autónoma de México, Apdo. Postal 20-364, Ciudad de México, México.
    }

\titlerunning{The lack of PNG response of bias-zero tracers}
\authorrunning{Celia Merino, Santiago Avila, et al.}

   \date{Received \today}

  \abstract{Constraining primordial non-Gaussianities (PNG) is one of the main goals of new-generation large-scale galaxy surveys. It had been proposed that unclustered tracers (with bias $b_1=0$) could be optimal for PNG studies, and that these could be found by selecting galaxies in bins of their local density. Here, we test this hypothesis in state-of-the-art simulations from the PNG-UNITsim suite with local $f_{\rm NL}=100$ and $f_{\rm NL}=-20$. We consider different parent tracer catalogues: all halos together, halos in large mass bins, and HOD models for LRGs and QSO. We then classify these tracers by their local density ($\delta_{t,R}$) and measure the linear bias ($b_1$) and PNG-response ($b_\phi$). Most $\delta_{t,R}$ bins show a PNG-response compatible with $b_\phi=0$ for all halos or the low-mass bin (log$M<11$). For high-mass halos (log$M>$12), QSO or LRG, we recover a trend closer to the universality relation ($b_\phi = 2 \delta_{\rm crit}(b_1-1)$) for $b_1>1$, but the $b_\phi(b_1)$ curve flattens to 0 below $\vert b_1\lvert<1$. Hence, we find $b_\phi\approx0$ for all bias-zero tracers considered. 
  The complex $\delta_{t,R}$-based selection causes their clustering to strongly deviate from simple assumptions, namely the universality relation and Poisson shot noise, hindering their capability to constrain PNG. 
 }


   \maketitle

   \nolinenumbers


\section{Introduction}

Cosmic inflation -- a phase of accelerated expansion during the first moments of the Universe --  was postulated as a solution to some of the problems of the Hot Big Bang theory associated with the initial conditions (ICs) \citep{Guth,Linde}. It also explains the origin of the inhomogeneities giving rise to the large-scale structure of the Universe (LSS). Although inflation is assumed to be part of the standard model in most modern cosmological analyses, the exact physics driving it remains unknown, as a wide variety of physical models can produce inflation. 

One observable that discriminates among families of models is primordial non-Gaussianities (PNG). 
We focus on the simplest PNG, local-\fnl \citep{Komatsu_2001}, which introduces quadratic expansion of a Gaussian gravitational field, $\phi_G$,
\begin{equation}
    \phi({\bf x}) = \phi_G({\bf x}) + \fnleqn \left( \phi_G({\bf x})^2 - \langle\phi_G({\bf x})^2\rangle\right),
\end{equation}
where $\langle \cdot \rangle$ represents a spatial average. Typically, this parameter is predicted to be \fnl$>\mathcal{O}(1)$ for inflation driven by multiple fields \citep{Byrnes_2010, Alvrez2014}. 

Although the primary effect of local-\fnl is to generate a primordial bispectrum, the non-linear evolution of the LSS allows us to study PNG through 2-point statistics. Galaxies ($g$) or halos ($h$) are biased tracers ($t$) of the matter ($m$) density field. At large scales, this is given by  a linear relation of overdensity fields ($\delta$): 
\begin{equation}
    \delta_t = b\cdot\delta_m.
\end{equation}
For Gaussian initial conditions, the large scale bias is simply a constant, $b_1$, the linear bias. However, when including PNG, the bias becomes scale-dependent \citep{Dalal_2008,Slosar_2008,Matarrese_2008}: 
\begin{equation}
    b(k,z) = b_1(z) + \frac{ \fnleqn\ b_\phi(z)}{\alpha(k,z)} \, .
    \label{eq:bk}
\end{equation}
Here, $b_\phi$ is a bias parameter associated with the gravitational field $\phi$, related with the overdensity field via 
\begin{equation}
 \begin{array}{l}
    \delta(k,z) = \alpha(k,z) \, \phi(k,z)\; ,  \\
    \; \;  {\rm with}\; \;  \alpha(k,z)=
     k^{2}\,T(k)\cdot
    \frac{2\,D(z)}{3\,\Omega_{M}}\,\frac{c^{2}}{H_{0}^{2}}\, ,
    \end{array}      
    \label{eq:delta_2_phi}
\end{equation}
where $D(z)$ is the growth factor normalised to $(1+z)^{-1}$ in the radiation era, $T(k)$ is the transfer function (with $T(k\to 0)\to1$), $\Omega_{M}$ is the current matter density, $H_0$ the Hubble constant, and $c$ the speed of light. The PNG universality relation is given by \citep{Dalal_2008}
\begin{equation}
    b_\phi = 2 \delta_{\rm crit}(b_1-p)\;\; {\rm with}\;\; p=1 
    \label{eq:universality}
\end{equation}
where $\delta_{\rm crit}$ is the critical density for spherical collapse ($1.686$).

Many galaxy surveys such as DESI, Euclid or SPHEREx, aim to constrain \fnl with \autoref{eq:bk}. 
Identifying observables that maximise the PNG signal would be highly valuable to better understand inflation.
In that context, \citet{castorina} postulated that {\it unclustered} ($b_1=0$) tracers would be optimal in a regime with negligible shot noise ($P_{ \rm shot}$), with an error given by
\begin{equation}
    \sigma(\fnleqn)^{-2 } \propto \frac{b_1^{\ 2}  b_\phi^{\ 2}}{\left( b_1^{\ 2}P_{mm}(k)+ P_{\rm shot}\right)^2} \, ,
    \label{eq:castorina}
\end{equation}
with $P_{mm} (k)$ the matter power spectrum.

The way proposed in that work to find bias-zero tracers is by binning tracers according to their local density.
They showed that unclustered tracers can be found in dark matter halo catalogues from cosmological $N$-body simulations with Gaussian ICs.  

The goal of this letter is to verify or falsify the feasibility of using bias-zero tracers selected by local overdensity to set strong constraints on PNG. For that, we use state-of-the-art $N$-Body simulations with PNG from the PNG-UNITsim suites \citep{PNG-unitsim, PNG-unitsim-XL}.

\section{Simulations}
\label{sec:sims}
The PNG-UNIT simulation is an $N$-Body simulation \citep{PNG-unitsim}, which is a twin of  one of the UNITsims \citep{unitsim}, with the same size ($L=1h^{-1}$Gpc), resolution ($N_{\rm part}=4096^3$, $m_p=1.2\times10^{9}\; h^{-1}M_\odot$), using the same white noise for the ICs, and \lcdm parameters (based on \citet{Planck15}), with the only difference of including a PNG of local-\fnl= 100. This makes it one of the PNG simulations with the largest mass resolution and one of the largest in terms of particles. In this work, we use the $z=1$ \textsc{Rockstar} \citep{rockstar} halo catalogues. These simulations are publicly available\footnote{\url{http://www.unitsims.org/} and  \url{https://opendata.pic.es/unitsims}}.

The PNG-UNITsim-XL we use here is a $L=3h^{-1}$Gpc simulation with $N=4096^3$ particles and initial conditions with \fnl$=-20$, which is a twin of a \fnl= 0 UNITsim-XL simulation. These simulations are fully described in \cite{PNG-unitsim-XL}. Both these simulations have been populated with a Halo Occupation Distribution (HOD) model \citep{Avila2020} matching the galaxy bias of DESI's LRG (using a  snapshot at $z=0.74$) and QSO (using $z=1.83$). The HOD is first set to have a maximum of $\sim1$ for central galaxies and is later downsampled (by a factor of $\sim 2$ and $\sim 40$ for LRG and QSO, respectively) to match the number density of DESI. In this work, we will show the results for the full density samples, while we checked that results are consistent --but noisier-- for the realistic-density sample.

\section{Methodology}
\label{sec:methods}

\begin{figure}
  \centering
  \includegraphics[trim={0 20 0 15}, width=0.98\linewidth]{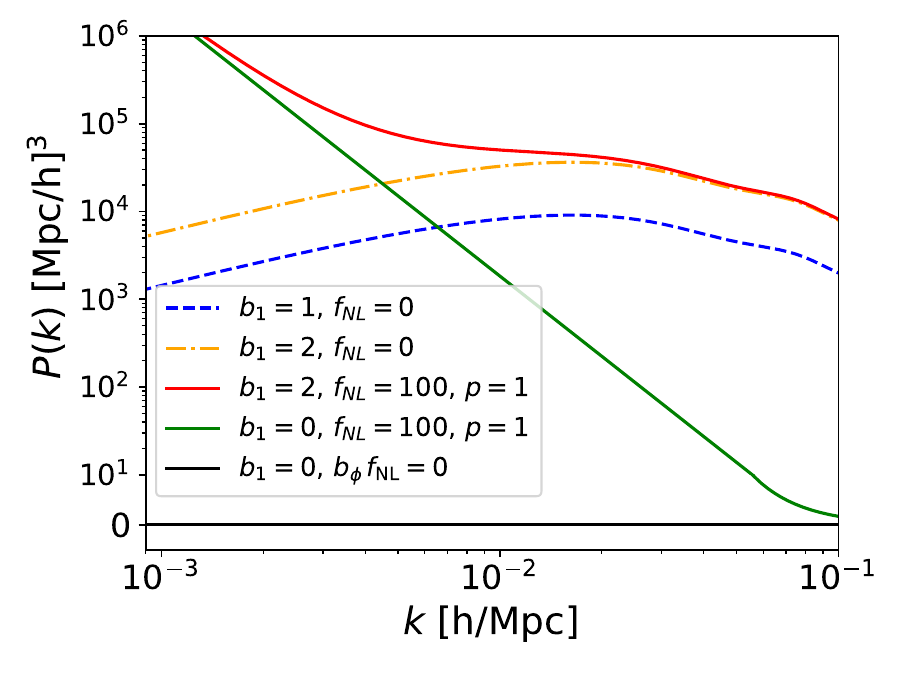}
     \caption{Theoretical auto-power spectra for matter ($b = 1$, \fnl= 0), a biased  tracer in a Gaussian Universe ($b_g=2$, \fnl$=0$), a biased universal tracer with PNG ($b_g=2$, \fnl$=100$, $p=1$), an unclustered but universal tracer ($b_g=0$, \fnl$=100$, $p=1$), and an unclustered and non-PNG-responsive tracer ($b_g=0$, $b_\phi=0$). 
     }
      \label{fig:motivation}
  \end{figure}

The potential of bias-zero tracers is illustrated in \autoref{fig:motivation}: an unclustered tracer in a Gaussian universe ($b_1=0$, black line) would cluster in PNG universes (green line) with a $1/k^2$ pattern. 
In this section, we describe the methodology to test that.  

\subsection{Density classification}
\label{sec:density_classification}

 Unclustered tracers are expected to be found in regions that are neither very dense or very empty. 
 We follow the approach of \cite{castorina} 
 and classify tracers according to their density in spheres of radius $R$, $\delta_{t,R}$, using by default $R=8 h^{-1}$Mpc.
 For this purpose, we use the \textsc{SciPy} function \textsc{cKDTree}. 

 We consider different options for parent samples: 
\begin{itemize}
    \item {\it all halos} in the PNG-UNITsims above 20 particles;
    \item PNG-UNItsim halos split in 3 {\it mass bins}: $10.5<{\rm log}M<11$, $11<{\rm log}M<12$ and ${\rm log}M>12$;
    \item {\it LRG} and {\it QSO} from an HOD applied to PNG-UNITsim-XL.
\end{itemize}


In this paper we focus on equal-number $\delta_{t,R}$ bins, starting from the parent catalogues described above. We use 20 bins for halos and 10 bins for galaxies.


For better comparison with \citet{castorina}, we also reproduced their method based on only splitting the parent sample into two local density subsamples. We then keep the lower one: $\delta_{t,R}<\delta_{t,R}^{\rm max}$. As we vary the parameter $\delta_{t,R}^{\rm max}$, we measure the linear bias of the sample, until we find $b_1\approx0$.

\subsection{Parameter fitting}

For each sub-sample of interest, we compute the tracer auto-power spectrum $P_{tt}(k)$ and the tracer-matter cross power spectrum $P_{tm}(k)$ with \textsc{nbodykit} \citep{nbodykit}. We additionally compute the matter power spectrum $P_{mm}(k)$, so that we can estimate the bias parameters by fitting the following equations: 
\begin{equation}
 \begin{array}{l}
    P_{tm}(k) = \left(b_1 + f_{\rm NL}\ b_\phi\ \alpha(k)^{-1}\right) P_{mm}(k)\, , \\
    P_{tt}(k) = \left(b_1 + f_{\rm NL}\ b_\phi\ \alpha(k)^{-1}\right)^2 P_{mm}(k) + \frac{A_{\rm sn}}{n}\, .
         \end{array}
\end{equation}
Where we fit the free parameters $b_1$, $b_\phi$, and $A_{\rm sn}$. Here, $A_{\rm sn}$ represents the shot noise deviation from the Poisson prediction ($1/n$). 

We use a Gaussian diagonal covariance \citep{cov}:
\begin{equation}
 \begin{array}{l}
    \sigma_{tt}(k)^2 = \left( P_{tt}(k) + \frac{A_{sn}}{n}\right)^2\cdot \frac{2\ (2\pi)^3}{4\pi\ k^2 \Delta k\ V}\;\; {\rm and}\\
    \sigma_{tm}(k)^2 = \left[ \left( P_{tt}(k) + \frac{A_{sn}}{n}\right)  P_{mm}(k) +  P_{tm}(k) ^2 \right] \frac{(2\pi)^3}{4\pi\ k^2 \Delta k\ V} , \\
     \end{array}
\end{equation}
where $V=L^3$ is the volume and $\Delta k$ is the $k$-bin width, which we set to the size of fundamental mode $k_f=2\pi/L$. 

We fit the parameters using an MCMC with the \textsc{emcee} library from the cross-power spectrum up to $k_{\rm max}=0.08h$Mpc$^{-1}$. We validate our methodology in \autoref{app:no_signal}.

\begin{figure}
    \centering
    \includegraphics[trim={0 20 0 15},width=0.98\linewidth]{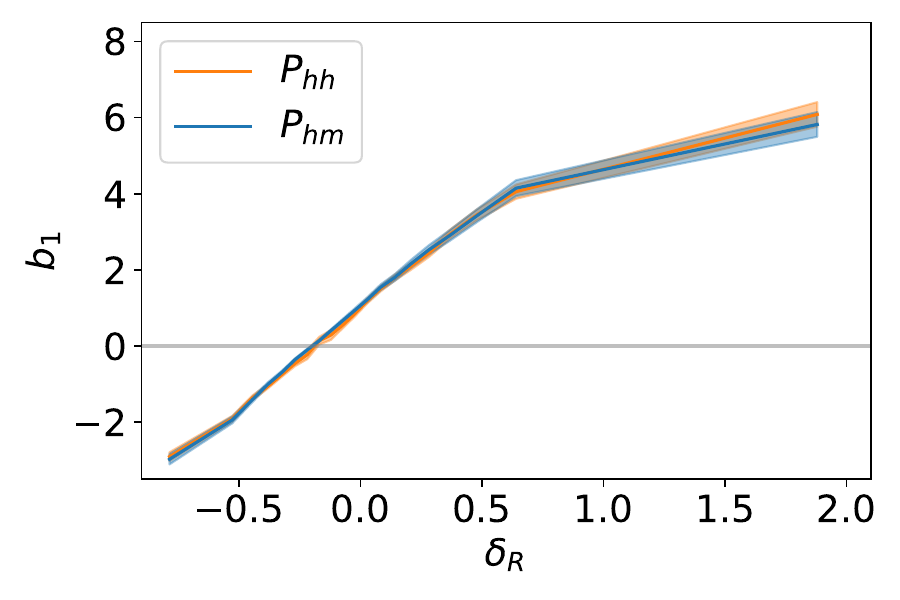}
    \caption{Measured linear halo bias $b_1$ (68\% c.l.) as a function of halo overdensity $\delta_{h,R}$ in spheres of $R=8 h^{-1}$Mpc for the  UNIT simulation (\fnl=0). Similar results are found for \fnl$=100$.
    }
    \label{fig:bias}
\end{figure}

\section{Results}
\label{sec:results}

\subsection{Halo density bins}
\label{sec:all_halos}

\subsubsection{Linear bias as a function of density}
\label{sec:all_halos_bias}

The linear bias of haloes grows monotonically with their local density $\delta_{h,R}$. This is shown for the UNIT simulation (\fnl= 0) haloes with at least 20 particles ($M_h>2\times10^{10}$) in \autoref{fig:bias}. For this measurement, we fit the linear bias ($b_1$) and shot-noise parameter ($A_{sn}$), while fixing \fnl= 0. \autoref{fig:bias} also shows the agreement between the results using the auto- and cross-power spectra. These results are in line with \cite{castorina}, and allow us to find the $b_1\approx0$ tracers. 

For the PNG-UNIT simulation (\fnl= 100) haloes, we additionally fit for $b_\phi$, while fixing \fnl= 100. The $b_1(\delta_{h,R})$ relation measured from the cross-correlation is nearly identical to the \fnl= 0 case. However, for the auto-correlation, the fits near $b_1\approx0$ are more numerically unstable, with $b_1$ absorbing the shot noise. For simplicity, we focus here on the cross-correlation results.

At linear scales, $\delta_h = b_1 \delta_M$. Hence, one would expect that we would measure something compatible with $b_1=0$ whenever $\delta_h=0$. However, this is not exactly what we see in \autoref{fig:bias}, where the curve does not cross $\{b_1=0, \delta_{h,R}=0\}$. 
This arises because $R=8 h^{-1}$Mpc is mildly non-linear; we verified that increasing R does not change our conclusions, while approaching the $\{b_1=0, \delta_{h,R}=0\}$ crossing.

Another interesting feature is that the best fit for $A_{sn}$ typically ranges in $20-40$, with $A_{sn}=27$ for the bin closest to $b_1=0$. \cite{castorina} already showed that these tracers were more stochastic than Poisson, but the factor found here is even higher, which might be due to having larger number density (and smaller baseline shot noise). 
This contribution of a factor $\sim 30$ in the increase of shot noise already suggests that using these tracers as optimal for PNG will be challenging, according to \autoref{eq:castorina}. We observe that $A_{sn}$ becomes larger for larger $R$, suggesting this effect comes from the exclusion of tracers.

\subsubsection{$b_\phi$ as a function of $b_1$}
\label{sec:allhalos}

We now move to the central idea of the paper using the PNG-UNITsim halos with \fnl= 100. On the top panel of \autoref{fig:bphi_all} we represent the fitted $b_\phi$ against the linear bias for {\it all halos} when classified in bins of $\delta_{t,R}$ as explained in \autoref{sec:density_classification}. 

We find that $b_\phi$ strongly deviates from the universality relation (\autoref{eq:universality}) marked by the dot-dashed line in \autoref{fig:bphi_all}. Moreover, we find that $b_\phi$ is compatible with zero for most of the $\delta_{h,R}$-bins. In particular, the bins close to $b_1=0$ are clearly vanishing. 

\subsection{Mass and halo density bins}
\label{sec:mass_bins}

\begin{figure}
    \centering
    \includegraphics[trim={0 20 0 15},width=\linewidth]{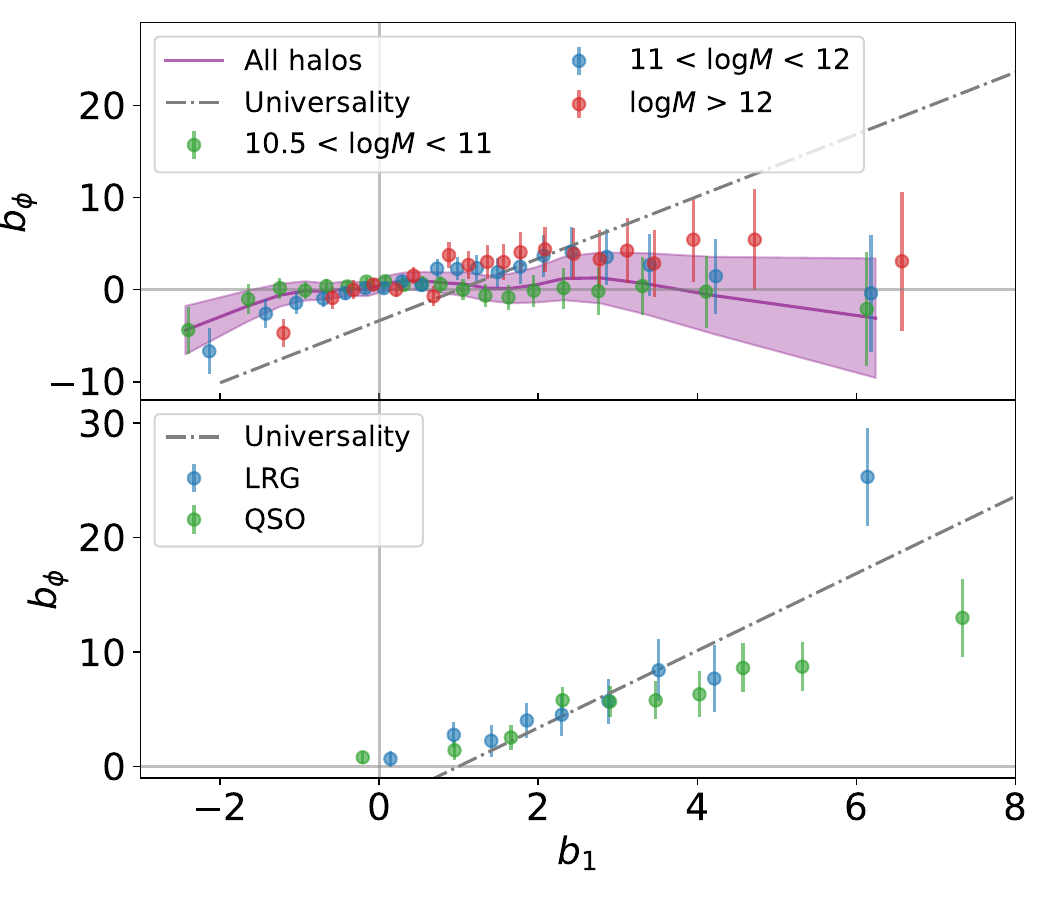}
    \caption{We show the PNG-response ($b_\phi$) as a function of the measured bias ($b_1$) for the tracers binned in their local overdensity. 
    {\bf Top:} Results for {\it all halos} (purple band) above $M=2\times10^{11}$ together of the PNG-UNITsim, based on the matter-halo cross-power spectrum. The black dot-dashed line shows the universality relation. We find that the PNG-response is well-below the universality relation and compatible with no response ($b_\phi\sim0$) for all the $\delta_R$-bins. We also show (as markers with errorbars) the results using mass bins as parent catalogues. {\bf Bottom:} Same results but for the LRG and QSO from PNG-UNITsim-XL. }
    \label{fig:bphi_all}
\end{figure}

A more realistic scenario is that you have a galaxy sample that is typically associated with a mass range. Indeed, in  \cite{castorina}, they test three mass bins as parent catalogues, similar to the three bins used here and defined in \autoref{sec:density_classification}.
We also show the $b_\phi(b_1)$ results for the mass bins in the top panel of \autoref{fig:bphi_all}. 
Remarkably, for the lower mass bin, we find a very flat $b_\phi(b_1)\sim0$ curve (as in \autoref{sec:all_halos}). On the other hand, the medium and high mass bins show a mild PNG response for $\lvert b_1\lvert>1$, but are still typically below the universality relation. 
The excess of shot noise becomes more moderate for mass bins, in increasing order, we find $A_{\rm sn}=$ 13, 7 and 2. 

\subsection{Galaxies in density bins}

Finally, we now use a more realistic scenario, where we sample the halo catalogue with an HOD, representing DESI QSO and LRG in the PNG-UNITsim-XL simulations. While we show the results for the high-density samples (see \autoref{sec:sims}), these are consistent with those from the realistic scenario, but less noisy. 

The results are shown at the bottom of \autoref{fig:bphi_all}. In this case, results follow a trend which is more similar to the universality relation. Nevertheless, the points near $b_1=0$ are still very compatible with $\lvert b_\phi\lvert\; =0$ and clearly below the universality relation. In this case, the excess of shot noise is more moderated with $A_{sn}=1.26$ for LRG and $A_{sn}=1.21$ for QSO. 

\subsection{Density splits for all parent catalogues}

\begin{figure}
    \centering
    \includegraphics[trim={0 20 0 15}, width=\linewidth]{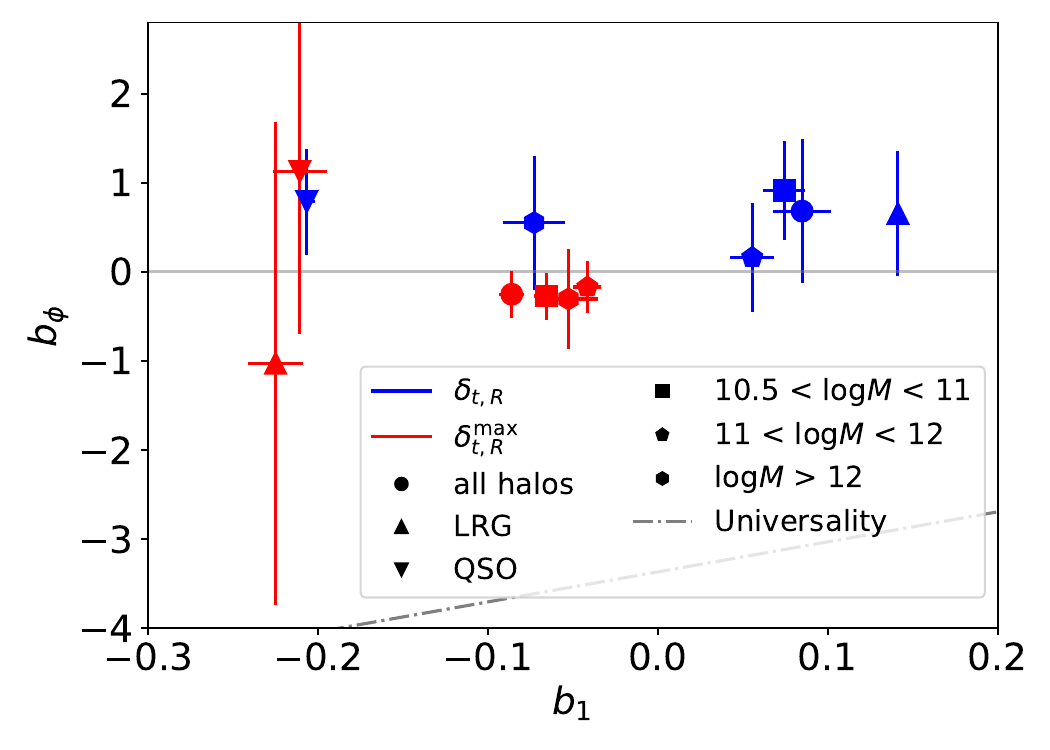}
    \caption{PNG response ($b_\phi$) and linear bias ($b_1$) measurements for the closest case to $b_1=0$ for the density bins (blue) and density split (red) samples. We consider different parent catalogues: all halos, halos in mass bins, and HODs (\autoref{sec:density_classification}). For all cases, the PNG response is clearly smaller than predicted by universality (dot-dashed line) and compatible with $b_\phi=0$.}
    \label{fig:density_split_bphi}
\end{figure}

Finally, we use an approach that is  matched to the original \cite{castorina}
proposal: splitting each of the parent halos in just 2 density bins, according to a $\delta_{r,R}^{\rm max}$. For each of the parent catalogues, we identify the $\delta_{t,R}^{\rm max}$ that makes $b_1$ closest to zero, and plot the values of $b_\phi$ for each of them in \autoref{fig:density_split_bphi}, and compare to the equivalent results from the $\delta_{t,R}$ bins.  
We find all of these cases to also be compatible with $b_\phi=0$. 

\section{Discussion and Conclusions}
In this work we have used state-of-the-art PNG simulations from the PNG-UNITsim suite to evaluate the efficiency of unclustered tracers ($b_1=0$) for the study of local-PNG as postulated by \cite{castorina}. 
We classified a parent tracer catalogue (all halos, halos in mass bins or HOD-based galaxies) by their local density $\delta_{t,R}$ in spheres of $R=8 h^{-1}$Mpc to measure their linear bias ($b_1$) and the PNG response ($b_\phi$) from the cross-power spectrum of the $\delta_{t,R}$-selected tracers and the matter field. 

When considering all halos at once or low-mass halos (10.5<log$M$<11), this method leads to 
$b_\phi(b_1)\approx0$ for all $\delta_{t,R}$ bins (except below $b_1\sim-1.5$). This implies that these tracers will
remain unclustered after the inclusion of PNG. More graphically, they will follow the black line of \autoref{fig:motivation} and not the green one. 

In the case of the medium and higher mass bins (11<log$M$<12, log$M>12$), we recover a slight hint for $b_\phi \neq0$, but only for $\lvert b_1\lvert>1$. 
When considering a parent sample of LRGs and QSOs from an HOD model, matching the galaxy bias of DESI, we recover a $b_\phi(b_1)$ curve closer to universality (\autoref{eq:universality}) for $b_1>1$, but that still flattens to $b_\phi=0$ for $b_1\approx0$. 

In summary, all unclustered tracers ($b_1\sim0$) found here remain unclustered with the inclusion of local-PNG (\autoref{fig:density_split_bphi}), strongly deviating from the universality relation. In addition to the density bins, we also consider splitting the sample into two sub-samples given a $\delta_{t,R}^{\rm max}$, finding qualitatively similar results.
We also verified that the results mentioned above are robust to scale-cuts, the usage of auto- and cross- power spectrum and the variation of the smoothing scale $R$ (from 4 to 30 $h^{-1}$Mpc).
A similar trend ($b_\phi(b_1=0)=0$) is also found in \autoref{app:matter}, when classifying halos in matter density instead of halo/galaxy density. 
We note that for $b_\phi=0$ the Fisher information in \fnl (\autoref{eq:castorina}) will vanish, leaving \fnl unconstrained. 
In addition, we find that these samples tend to have super-Poissonian shot noise, sometimes by a factor $\sim30$, also complicating the optimality. 

This $b_\phi(b_1\approx0)\approx0$ relation reduces the effectiveness to constrain PNG. As an example, if we do a Fisher forecast for the LRG and QSO modelled in PNG-UNITsim-XL for the DESI volume and number density, we obtain, respectively, $\sigma_{f_{\rm NL}}=4.1$ and $\sigma_{f_{\rm NL}}=3.8$. When considering their $\delta_{h,R}$-bin closest to $b_1=0$, even if neglecting shot-noise, we obtain respectively, $\sigma_{f_{\rm NL}}=133$ and $\sigma_{f_{\rm NL}}=102$.
Additionally, the $b_\phi(b_1=0)=0$ relation would also hinder a multi-tracer analysis, where $\sigma(f_{\rm NL})^{-2}\propto\lvert b_1^A  b_\phi^B - b_1^B b_\phi^A\lvert$ \citep{Barreira_multitracer}.

The lack of a PNG response appears in tension with the results reported by \citet{castorina}, where they found a PNG response of up to $b_\phi\approx 2.5$ for a $b_1=0$ tracer from their log$M$>12.5 parent catalogue\footnote{We verified our conclusions remain unchanged for log$M$>12.5.}.
This difference could possibly arise because that work used separate-universe simulations (with \fnl= 0) to estimate $b_\phi$ as a derivative of the number of tracers with respect to the amplitude of primordial fluctuations ($A_s$). Whereas the separate-universe approach has been tested in halo mass bins \citep{Biagetti}, the selection based on the local halo density inherently brings higher order terms (in density and bias) to the selection that could in turn have cosmology dependency (including $A_s$). 
This possibility is supported by \citet{PNG_density_split} who found that $\delta_{h,R}$-selected tracers (from a parent catalogue of log$M>13.5$ halos) have a PNG response that deviates from the separate-universe prediction (see their Figure 9). 

Exploring the PNG signal from alternative methods and samples is a powerful way to maximise the information from current and future missions. This work remarks on the importance of developing in parallel realistic suite of simulations to validate and improve the modelling of these less standard statistics. 

\begin{acknowledgements}
      We thank discussions with E. Paillas, J. L. Bernal, R. Angulo and I. Sevilla. 
      SA has been funded by MCIN/AEI/10.13039/501100011033 and FSE+ (Europe) under project PID2024-156844NA-C22 and the RYC2022-037311-I fellowship. VGP is supported by the CNS2024-154242 grant. 
      {\it Contributions:} CM led the analysis and figure making. SA led the conceptual design of the project and writing of the paper. AGA contributed to the analysis and interpretation. AA led a first version of the analysis. AGA, SA and VGP led the simulation construction. JMR constructed the HOD catalogues. 
      All the authors contributed to the manuscript.
\end{acknowledgements}

\bibliographystyle{aa}
\bibliography{biblio_list}

\appendix
\section{The lack of PNG signal}
\label{app:no_signal}

The flat curve for $b_\phi(b_1)=0$ at the top of \autoref{fig:bphi_all} can be somewhat surprising at first. To discard errors in the analysis, we verified that for simple halo-mass bins we recover a $b_\phi(b_1)$ law which is very close to the universality relation (\autoref{eq:universality}). In fact, we are able to reproduce the results from \cite{PNG-unitsim} with the analysis tools used for this letter. 

\begin{figure*}
    \centering
    \includegraphics[width=0.5\linewidth]{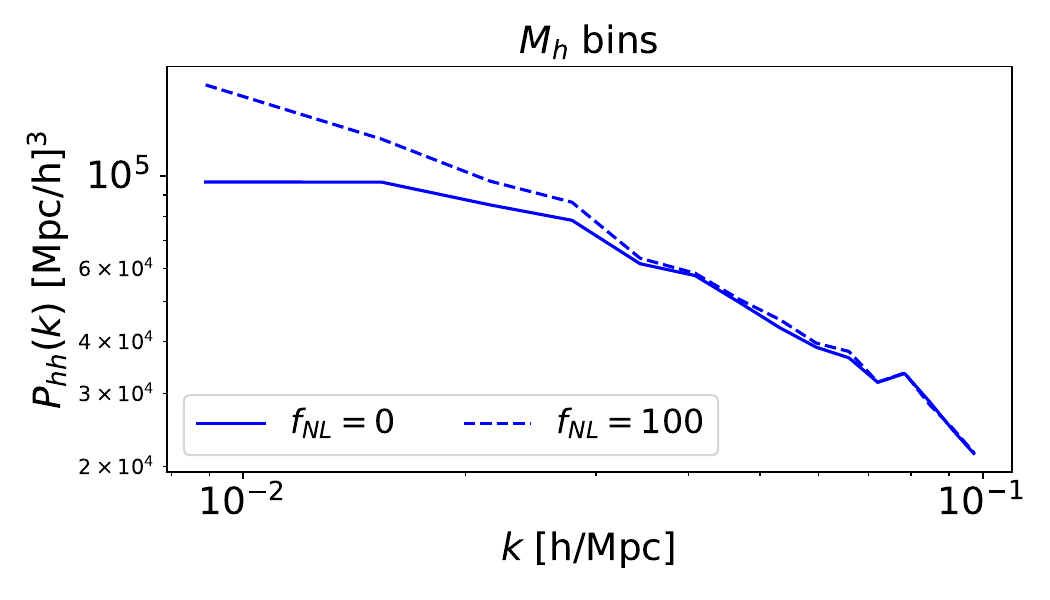}\includegraphics[width=0.5\linewidth]{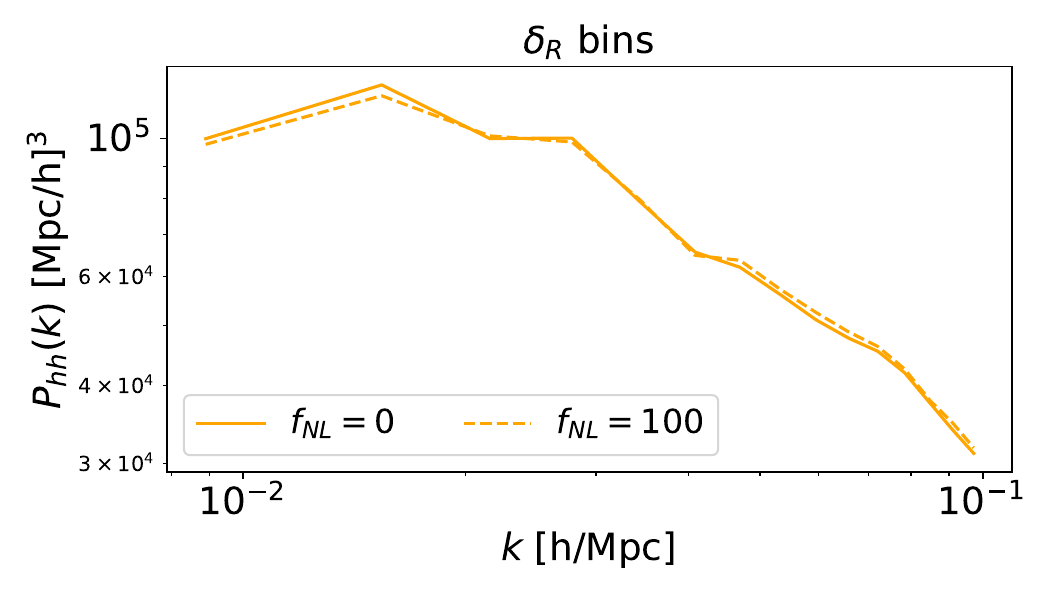}
    \caption{Power spectrum of a bin with $b_1\sim 3.2$ selected by halo mass ($M_h$, {\bf left}) and by local halo overdensity ($\delta_{h,R}$, {\bf right}). Whereas the effect of introducing PNG (\fnl$=100$) has a strong effect on mass-selected bins, the $\delta_{h,R}$-bins show near identical clustering for \fnl= 0 and \fnl= 100.}
    \label{fig:mass_vs_density_bins_pk}
\end{figure*}

Another way to verify the tools used here is to select a bin in mass with bias $b_1 \sim3$, and compare it to one of the bins in $\delta_{h,R}$ with a similar bias value (using all halos as parent catalogue). This is shown in \autoref{fig:mass_vs_density_bins_pk}, where we observe a very clear PNG effect at low-$k$ for the mass bins, whereas we do not observe any PNG response for the density bins.

\section{Classifying by dark matter overdensity}
\label{app:matter}

In the main body, we focused on using biased tracers (halos or galaxies) for the estimation of local density, $\delta_{t,R}$. This is motivated by assuming this is what we can do with galaxy surveys. One could think of other ways of estimating local density, such as lensing (either from CMB or from galaxy shear), which will be a more direct probe of the matter field, although typically projected to a 2D sky over a long radial kernel. 

\begin{figure}
    \centering
    \includegraphics[width=\linewidth]{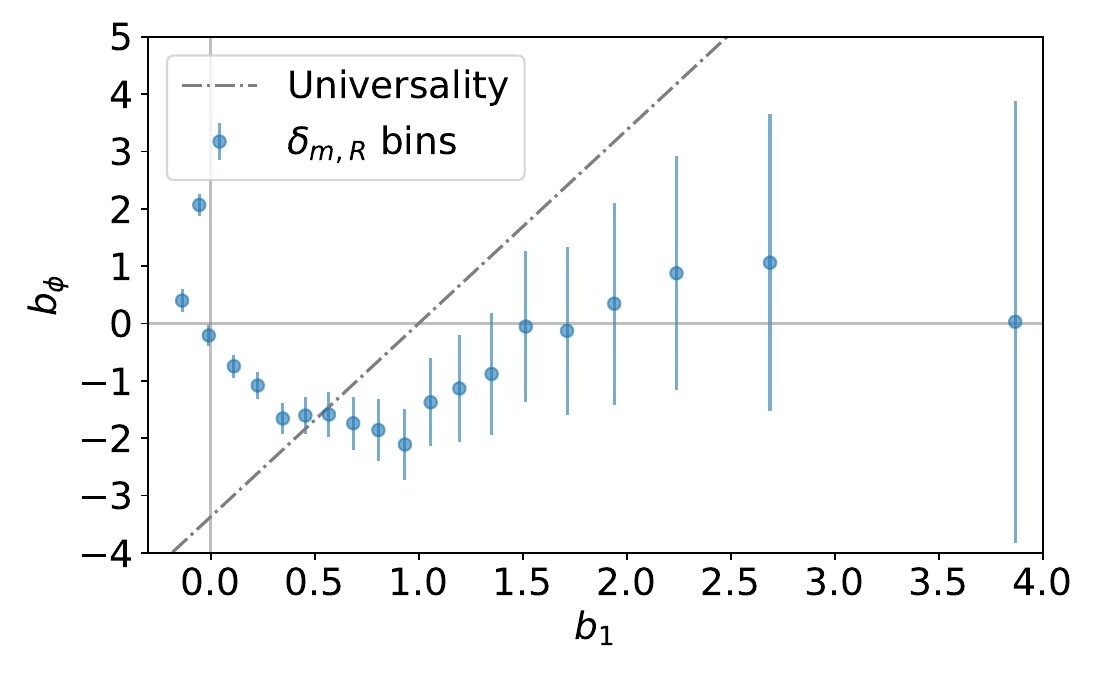}
    \caption{PNG response $b_\phi$ as a function of linear bias $b_1$ for halos classified as a function of local {\bf matter} density $\delta_{m,R}$ using all PNG-UNITsim (\fnl= 100) halos above 20 particles.}
    \label{fig:DM}
\end{figure}

For this reason, in this appendix we test an ideal scenario where we would have access to the matter field, $\delta_{m,R}$ and we can classify halos according to this density. From the resulting classification, we can compute the linear ($b_1$) and PNG ($b_\phi$) bias as before, obtaining \autoref{fig:DM}.
We find that there is a strong deviation from universality, with negative $b_\phi$ for $0.5<b_1<1.5$. Nevertheless, what remains unchanged with respect to other tracers is that for $b_1=0$ we still find a clear $b_\phi=0$.

\end{document}